\documentclass{webofc}
\usepackage[varg]{txfonts}   
\begin{document}
\title{Monitoring activities of satellite data processing services in real-time with SDDS Live Monitor}
%
%

\author{
  \firstname{Minh Duc}
  \lastname{Nguyen}
  \inst{1}
  \thanks
  {\email{nguyendmitri@gmail.com}}
}

\institute{Skobeltsyn Institute of Nuclear Physics, Lomonosov Moscow State University, Moscow, Russia}

\abstract{%
  This work describes Live Monitor, the monitoring subsystem of SDDS – an automated system for space experiment data processing, storage, and distribution created at SINP MSU. Live Monitor allows operators and developers of satellite data centers to identify errors occurred in data processing quickly and to prevent further consequences caused by the errors. All activities of the whole data processing cycle are illustrated via a web interface in real-time. Notification messages are delivered to responsible people via emails and Telegram messenger service. The flexible monitoring mechanism implemented in Live Monitor allows us to dynamically change and control events being shown on the web interface on our demands. Physicists, whose space weather analysis models are functioning upon satellite data provided by SDDS, can use the developed RESTful API to monitor their own events and deliver customized notification messages by their needs.
}
\maketitle
\section{Introduction}
\label{introduction}
The ultimate goal of research in space weather is creating complex operational magnetosphere-ionosphere models which allow predicting the radiation risk for space satellites in different orbits, and also estimating the occurrence risk of technological disasters due to magnetic storms and charged particle precipitations. Currently, solutions to the problem involve collecting data from all available satellites and ground stations, each of which measures a limited number of parameters of space weather, and generating forecasts based on the result of its analysis. Such task is challenging because data formats, storage methods, datasets and values of measured parameters differ between satellites. To achieve this goal an automated system called SDDS has been created at SINP MSU. SDDS automatically connects to different data storage servers of various satellites, downloads real-time data (both decoded txt-files and binary telemetry) whenever available, decodes binary telemetry, processes decoded data and stores it in a unified database.

Satellite data collected by SDDS is used as the primary source of input data for space weather operational models created at SINP MSU. Since the correctness and preciseness of space weather models depend much on the input data, it is critical to be sure that data is correctly processed. It is also important to detect errors in any step of the data processing cycle as soon as possible so that they could be fixed quickly and their consequences could be prevented in the future. Solution to the problem is monitoring all activities of all components involved in data processing and sending alerts to responsible people (satellite developers, system administrators, workers on duty, etc.) so they could take measures in time.
This paper is organized as follows: in the second section, we consider several existing solutions to the problem compared to our approach; in section 3, we give a more detailed view of the overall architecture of the Live Monitor subsystem. A brief description of the backend library is considered in section 4. Also in this section, we explain how events are classified and how alerts are sent via different mechanisms. Section 5 is dedicated to event representation on the web interface. In section 6, we give a brief description of using Live Monitor’s API to create a customized monitoring service. In conclusion, we give a short resume of our completed work and describe our vision of the future perspective.

\section{Related works}
\label{related_works}
Since SDDS is working on Linux, the first solution came to mind is using an existing open source monitoring solution such as Zabbix \cite{zabbix}, Nagios \cite{nagios}, or MMonit \cite{mmonit} to accomplish the task. But the detailed functional analysis showed that these systems were mainly designed for monitoring IT infrastructure (servers, routers, switches, etc.), system and network services. Zabbix and Nagios support application monitoring but this function is commercial. To monitor a custom service (or application) both Zabbix and Nagios assume that one must write a wrapper which runs a number of tests to check the service and produces standardized output data interpretable by the interface of the solution. The server component of Zabbix or Nagios then call this wrapper directly or via a client agent on a regular basis to check the service. Such solutions are not suitable for SDDS because of several reasons. Processing data of different satellites involves various components, the components can change dynamically, and working states of each component also differ. Writing a wrapper for each satellite would lead to a big amount of source code to be maintained. Additional checks on a regular basis for too many services would affect the overall performance of the operating system. Thus, a lightweight event-driven monitoring mechanism would be better.

A better solution is requiring all components of SDDS to inform about their current states in such a way that states could be treated uniformly. Each event of each state can be logged as a record to a journal file. When the log record is produced, we can use a message broadcasting server to either send the record to a web interface to show it to an operator or deliver the record via email or a messenger service directly to him. We have developed a library in three major programming languages Python, PHP, and JavaScript solely for this purpose.

While this approach can be applied perfectly to under development programs and internal components of SDDS being maintained by us, the same is not true for external programs that are used by SDDS to extract scientific data from the raw binary data received from satellites. It is a real challenge because most of these programs were written using different tools, both open source and commercial, and programming languages. The explanation of this fact is that data formats of different satellites differ and physicists use the tools they know best to achieve their goals. Changing these programs is not reasonable, mainly because either many of them were written such a long time ago that no one knows exactly what was implemented inside, or they are such complicated, so changes can lead to unexpected behaviors, resulting in erroneous data. To overcome this difficulty, we have decided to run these programs inside a wrapper which uses the developed library.
Our final solution became the Live Monitor subsystem that is used to monitor all activities of SDDS in real-time. We have also developed a RESTful API so that physicists, whose space weather models are working based on the data supplied by SDDS, can use it to create their own monitoring scheme and deliver customized alerts to their customer.

\section{Live Monitor's architecture}
\label{live_monitor_architecture}
Live monitor subsystem consists of the following components: a backend logging library, RabbitMQ message broker, a RESTful API backend, a frontend UI library. An illustration of the architecture is shown below in figure~\ref{figure_architecture}.

\begin{figure}[ht]
\centering
\includegraphics[height=6.5cm, clip]{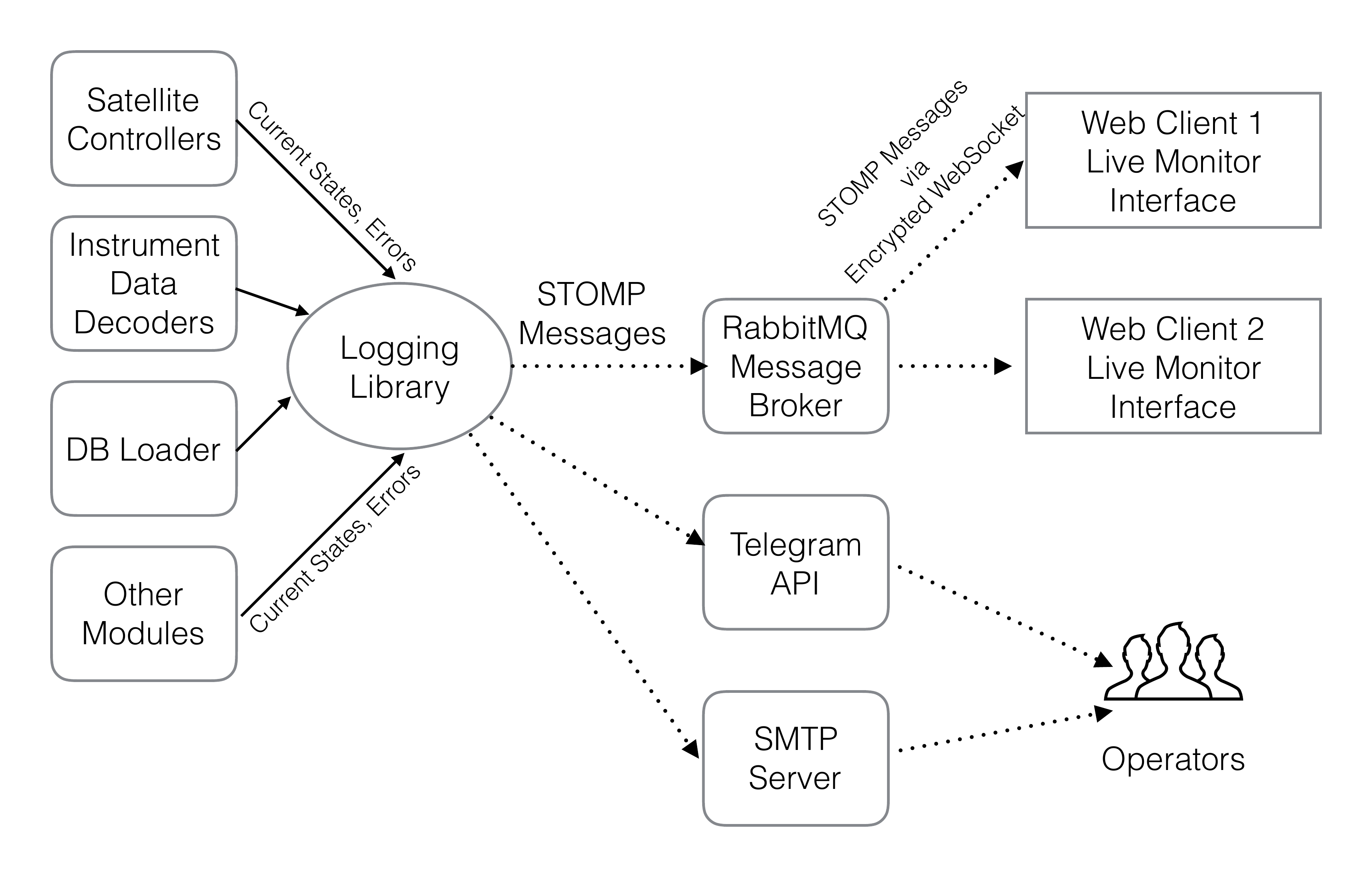}
\caption{Live Monitor's architecture}
\label{figure_architecture}
\end{figure}

The backend logging library is a customization of the popular Python Logging module. Every component of the SDDS system uses the library to inform about its states while running. External programs are executed within a wrapper of the SDDS global controller. When an external program crashes or returns an error code, the wrapper informs about it using the library. Besides the standard behavior which is writing short text messages in different log levels to a log file, the library sends these text messages to the RabbitMQ message broker [4] via a TCP socket. Messages are formatted using the STOMP protocol [5]. RabbitMQ in turn broadcast received text messages to all active web clients. When an error occurs during a certain state of data processing, the library informs operators about the error by sending a corresponding error description directly to the operators via the Telegram messenger service and/or email. Notification features can be turned on and off by editing proper configuration files.

Each data processing cycle is divided into stages. For example, a processing cycle of data from Meteor-M2 satellites consists of the following stages: connecting to data sources, downloading new raw data files, extracting scientific binary data files from raw ones, processing binary data files, adding processed data to the database, moving both raw and binary files to the local storage. Processing stages of different satellites differ. Stages to be displayed on the web interfaces of all components are customizable. One just needs to define a list of all states in the component configuration file; the library automatically does everything else. Since a satellite has a number of instruments on board, the whole processing stage consists of sub-stages each of which belongs to each instrument. An example is shown below in figure~\ref{figure_web_ui}.

\begin{figure}[ht]
\centering
\includegraphics[height=6.5cm, clip]{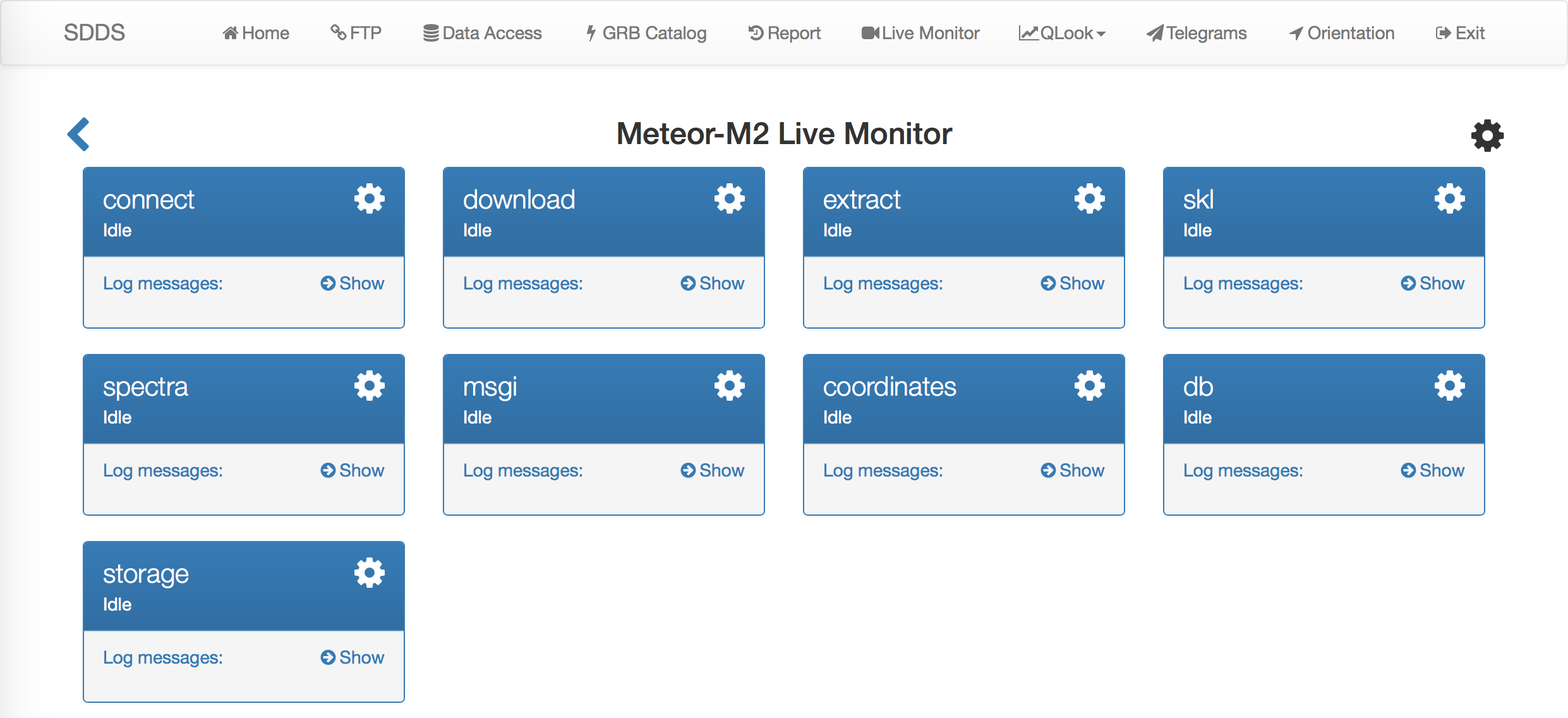}
\caption{Live Monitor's architecture}
\label{figure_web_ui}
\end{figure}

\section{Live Monitor's backend library}
\label{backend}
The primary goal of Live Monitor’s backend library is logging all events from all components of SDDS as text messages to log files. If an SDDS's component needs to be monitored in real-time, one can change the proper property in the component configuration file, and all logging messages will be passed to RabbitMQ message broker and then delivered to all web clients. To monitor external programs used for decoding scientific data from raw one, SDDS execute them in a wrapper. The wrapper logs all start and stop execution points and also error codes and output of these programs using the logging library.

The logging library supports four levels of logging messages: debug, info, warning, and error. Error messages are logged when an error occurs during the data processing cycle which could lead to incorrectly processed data or cause a component failure. Warning messages are logged for minor errors that do not affect the data correctness and normal functioning. Info messages are just normal text descriptions of events during the data processing cycle. Debug messages include diagnostic information that is helpful in failure investigation. If the Live Monitor property is enabled in the configuration file of a satellite, error messages will be passed directly to operators on duty using the Telegram messenger service.

Another responsibility of Live Monitor’s backend is to control what should be shown on the web interface during a data processing cycle of a satellite. When a satellite is created, the satellite operator defines in the configuration what stages the data processing cycle consists of and what instruments are working on the satellite. A RESTful API has been developed to deliver this kind of information to the frontend library flexibly and to allow remote work with Live Monitor subsystem.

RabbitMQ was used as the message broker server to deliver short messages to the frontend web client, primarily, because of its excellent documentation and community support. Also, RabbitMQ provides libraries for all major programming languages, which is a big advantage. In our case, the type of message distribution is “publish-subscribe” with no guaranteed delivery. In this scenario, RabbitMQ demonstrates an impressive performance and stability according to the performance benchmark in the study \cite{stomp_performance}. Also, RabbitMQ requires less effort in configuration settings and implementation, and it’s more reasonable compared with Apache ActiveMQ Apollo \cite{activemq} or ZeroMQ \cite{zeromq}.

\section{Live Monitor’s frontend library}
\label{frontend}
The main goal of the frontend library is to control how a data processing cycle of a satellite should be shown on the web interface. When a user opens the Live Monitor’s page of a satellite, the frontend library sends a request to the backend according to the RESTful API to retrieve necessary information of what should be shown. The answer from the backend is a JSON object that consists of a number of stages and a number of instruments decoders involved in the data decoding stage. After that, the frontend library establishes a WebSocket connection with the RabbitMQ message broker. When a text message of a stage arrives, the frontend library parses its content and changes the visual appearance of the stage. Below is an example of the common “connecting-to-data-sources” stage of each satellite.

\begin{figure}[ht]
\centering
\includegraphics[height=6.5cm, clip]{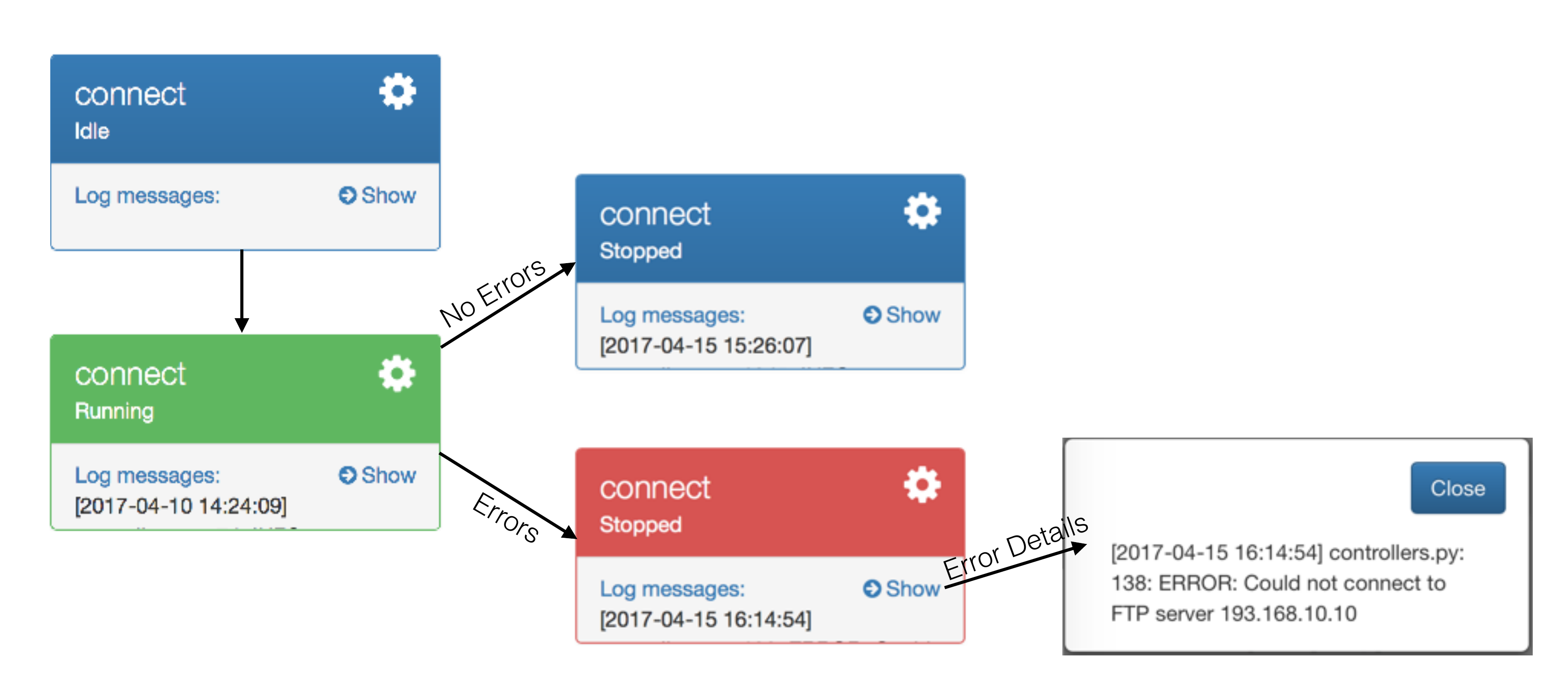}
\caption{Different states of the “connecting-to-data-source” stage of a satellite}
\label{figure_statuses}
\end{figure}

We use WebSocket protocol \cite{websocket} to deliver messages from RabbitMQ broker to subscribed web clients because of its performance. The big advantage of using WebSocket instead of AJAX long polling requests \cite{ajax} is that we can push notifications to clients when an event occurs through a bi-directional socket. It significantly reduces the workload of the server. In the case of using AJAX long polling requests, each web client would have to poll the broker regularly for new events. This approach leads to significant performance overheads, especially when we have a large number of clients, due to the unnecessary check data packets sending back and forth between the broker and the clients.

We use Simple (or Streaming) Text Oriented Messaging Protocol (STOMP) as the communication language between the broker and the clients. STOMP provides an interoperable wire format so that STOMP clients can communicate with any STOMP message broker to provide easy and widespread messaging interoperability among many languages, platforms, and brokers. It suits our needs best. STOMP is a very simple and easy to be implemented protocol, coming from the HTTP school of design. It is very easy to write a client to get yourself connected. For example, one can use Telnet to log into any STOMP broker and interact with it. We have managed to write three STOMP clients (in Python, PHP, JavaScript) in just a couple of hours and to integrate them to the RabbitMQ broker.

\section{Customized monitoring service based on Live Monitor}
\label{customized_service}
At the moment of writing, Live Monitor is used mostly to monitor data processing of satellites. However, it is also possible for one to create his own monitoring service based on Live Monitor’s API.

First of all, one must be a registered user of our SDDS system to be authorized to use Live Monitor’s API. To create a customized monitoring service, a user must send a POST request with a proper JSON object body which consists of a service name, all its stages, and component names involving in each stage. When such request is received, Live Monitor’s backend checks the request body for correctness. If the request is correct, the backend will create a RabbitMQ channel dedicated only to delivering messages of the requested service and a configuration file. As the answer, the user receives a JSON object which consists of a URL to the page of the requested service, a pair of login and password to be used in user's program to deliver messages to the RabbitMQ message broker. To control the behaviour of the service, one must send proper requests to the Live Monitor’s backend.
Currently, the following operations are supported by Live Monitor:
\begin{itemize}
\item create/delete a customized monitoring service;
\item switch a monitoring service on/off;
\item switch Telegram message delivery for a service on/off.
\end{itemize}

\section{Conclusion}
\label{conclusion}
SDDS is the primary satellite data processing system being used at SINP MSU. The Live Monitor subsystem of SDDS system has been deployed from the first day of operation. During 12 months of operation Live Monitor helped us react to any occurred event in data processing in time. Since all stages of data processing are shown in details, Live Monitor allows us to identify and localize the scope of a problem immediately and hence prevent or fix them quickly. In future, we plan to support more operations to control the behavior of monitoring services, such as manipulating a stage name or a component name of a stage, controlling the monitoring behavior of each stage and component of a service independently, and so on.

\section*{Acknowledgments}
We would like to thank Dr. Vladimir Kalegaev for helpful discussions on satellite data processing and clear problem statements. Also, we would like to thank Dr. Alexander Kryukov for his support in all aspects. This project is supported by RSF grant \#16-17-00098.
%
%
%

\end{document}